\documentstyle[12pt,a4wide,amssymb,epsfig]{article}
\begin{document}
\flushbottom
\renewcommand{\thefootnote}{\fnsymbol{footnote}}
\pagestyle{empty}
\setcounter{page}{0}
\begin{flushright}
DFTT 84/95
\\
hep-ph/9512349
\end{flushright}
\vspace*{1cm}
\begin{center}
\LARGE \bf
Possible tests for sterile
neutrinos\protect{\footnote{\normalsize Talk
presented by S.M. Bilenky at TAUP 95,
Toledo (Spain), September 17-21, 1995}}
\vspace*{1cm}
\\
\Large \mediumseries
S.M. Bilenky$^{\mathrm{a}}$
and
C. Giunti$^{\mathrm{b}}$
\\
\vspace{0.5cm}
\large \rm
\begin{tabular}{c}
$^{\mathrm{a}}$Joint Institute for Nuclear Research,
Dubna, Russia.
\\
$^{\mathrm{b}}$INFN and
Dipartimento di Fisica Teorica, Universit\`a di Torino,
\\
Via P. Giuria 1, 10125 Torino, Italy.
\end{tabular}
\\
\vspace*{1cm}
Abstract
\\
\vspace{0.5cm}
\normalsize
\begin{minipage}[t]{14cm}
It is shown that
the future SNO and Super-Kamiokande experiments,
in which high energy $^8\mathrm{B}$ neutrinos
will be detected through the observation of
CC, NC and $\nu$--$e$ elastic scattering processes,
could allow to reveal in a model independent way
the presence of sterile neutrinos
in the flux of solar neutrinos on the earth.
\end{minipage}
\end{center}

\newpage
\pagestyle{plain}
\setcounter{footnote}{0}
\renewcommand{\thefootnote}{\arabic{footnote}}

We discuss here some possibilities to
reveal transitions of solar $\nu_e$'s
into sterile states \cite{BG95}
in the future
SNO \cite{SNO}
and
Super-Kamiokande (S-K) \cite{SK}
experiments.
According to the hypothesis of neutrino mixing
\cite{PONTECORVO},
the flavor neutrino fields $\nu_{\ell L}$ are linear combinations
of the left-handed components of neutrino fields
$\nu_{iL}$
with masses $m_i$:
\begin{equation}
\nu_{\ell L}
=
\sum_{i}
U_{\ell i}
\nu_{iL}
\;,
\label{01}
\end{equation}
where $U$ is a unitary mixing matrix.

{}From the data of LEP experiments
it follows that
the number of flavor neutrinos
is equal to three.
If the total lepton number is conserved,
massive neutrinos are Dirac particles
and the number of massive neutrinos is equal to three.
On the other hand,
in the Majorana case
the number of light neutrinos with definite mass
can vary from three to six
(see \cite{BILENKYPETCOV87}).
If this number is more than three,
in addition to Eq.(\ref{01})
we have
\begin{equation}
\left( \nu_{\ell R} \right)^{c}
=
\sum_{i}
U_{\bar\ell i}
\nu_{iL}
\;,
\label{E02}
\end{equation}
where
$ \displaystyle
\left( \nu_{\ell R} \right)^{c}
=
{\cal{C}} \overline{\nu}_{\ell R}^{T}
$
is the charge-conjugated field.
Due to the mixings
(\ref{01}) and (\ref{E02}),
not only
transitions between active neutrinos
$ \nu_{\ell L} \to \nu_{\ell' L} $
are possible,
but also transitions between
active and sterile neutrinos
$ \nu_{\ell L} \to \bar\nu_{\ell'L} $
($ \nu_{\ell R} $ and $ \bar\nu_{\ell L} $
are the quanta of the right-handed fields
$ \nu_{\ell R} $).
The quanta of the right-handed fields
are called sterile
because they
do not enter in the Lagrangian
of the standard weak interactions.

Transitions of active neutrinos
into sterile states
can exist only in models beyond the Standard Model.
Thus,
the discovery of transitions of active neutrinos into sterile states
would have a fundamental importance
for the theory.
The existing data
do not exclude such transitions.
The data of all solar neutrino experiments
can be explained if $\nu_e$
is mixed with a sterile neutrino $\nu_s$.
Assuming the correctness of the Standard Solar Model
(see \cite{BAHCALL}),
the following values of the mixing parameters
were obtained
\cite{SOLSTE}:
\begin{equation}
\Delta m^2 \simeq 5 \times 10^{-6} \, \mathrm{eV}^2
\quad \mbox{and} \quad
\sin^2 2\theta \simeq 7 \times 10^{-3}
\;.
\label{E03}
\end{equation}
Let us remark that
these values of the mixing parameters
are compatible
with the constraints on
$\nu_e$--$\nu_s$
mixing obtained
\cite{BBN}
from big-bang nucleosynthesis.

Information about
the transitions of active neutrinos into sterile states
can be obtained
only through the measurement
of the total transition probability
of an active neutrino
$\nu_\ell$
into all possible active neutrino states,
$ \displaystyle
\sum_{\ell'=e,\mu,\tau} \mathrm{P}_{\nu_{\ell}\to\nu_{\ell'}}
$,
or any average of this quantity.
If it occurs that this total transition probability
is less than one,
it would mean that transitions
of active neutrinos into sterile states
take place.

It is obvious that we can obtain information
about the total transition probability
only through the investigation of neutral-current (NC)
neutrino processes.
In 1996 two new solar neutrino experiments
will start.
In the SNO experiment
\cite{SNO}
solar neutrinos will be detected through the observation
of the following
charged-current (CC),
elastic (CC and NC)
and NC processes:
\arraycolsep=0.4cm
\begin{eqnarray}
&
\nu_{e} + d \to e^{-} + p + p
&
\mathrm{(CC)}
\label{E:CC}
\\
&
\nu + e^{-} \to \nu + e^{-}
&
\mathrm{(ES)}
\label{E:ES}
\\
&
\nu + d \to \nu + p + n
&
\mathrm{(NC)}
\label{E:NC}
\end{eqnarray}
In the CC process (\ref{E:CC}),
the electron spectrum will be measured
and the flux
of solar $\nu_e$'s on the earth
as a function of neutrino energy $E$,
$\phi_{\nu_e}(E)$,
will be determined
\cite{SNO}.

In the S-K experiment
\cite{SK}
solar neutrinos will be detected
through the observation of the ES process (\ref{E:ES}).
The event rate is expected to be about 100 times
larger than in the current Kamiokande experiment
and the spectrum of the recoil electrons
will be measured with high accuracy.

In both the SNO and S-K experiments,
due to the high energy thresholds
($\simeq 6 \, \mathrm{MeV}$
for the CC process,
$2.2 \, \mathrm{MeV}$
for the NC process
and
$\simeq 5 \, \mathrm{MeV}$
for the ES process),
only neutrinos coming from
$^8\mathrm{B}$ decay
will be detected.
The energy spectrum of the initial $^8\mathrm{B}$ $\nu_e$'s
can be written as
\begin{equation}
\phi_{\mathrm{B}}(E)
=
\Phi_{\mathrm{B}}
\,
X(E)
\;.
\label{E500}
\end{equation}
Here $X(E)$ is a known
\cite{BAHCALL}
normalized function
determined by the phase space factor
of the decay
$ ^8\mathrm{B} \to \mbox{} ^8\mathrm{Be} + e^{+} + \nu_{e} $,
and
$\Phi_{\mathrm{B}}$
is the total flux of initial $^8\mathrm{B}$ solar $\nu_{e}$'s.

Let us consider first
the NC process (\ref{E:NC}).
The total NC event rate,
$ N^{\mathrm{NC}} $,
is given by
\arraycolsep=0cm
\begin{eqnarray}
N^{\mathrm{NC}}
\null & \null = \null & \null
\int_{E_{\mathrm{th}}} \hskip-3mm
\sigma_{{\nu}d}^{\mathrm{NC}}(E)
\begin{array}[t]{l}
\displaystyle \sum
\\
\scriptstyle \ell=e,\mu,\tau
\end{array}
\hskip-4mm
\mathrm{P}_{\nu_{e}\to\nu_{\ell}}(E)
X(E)
{\mathrm{d}} E
\,
\Phi_{\mathrm{B}}
\nonumber
\\
\null & \null = \null & \null
\left\langle
\begin{array}[t]{l}
\displaystyle \sum
\\
\scriptstyle \ell=e,\mu,\tau
\end{array}
\hskip-4mm
\mathrm{P}_{\nu_{e}\to\nu_{\ell}}
\right\rangle_{\hskip-2pt\raisebox{4pt}{\scriptsize{NC}}}
X_{{\nu}d}^{\mathrm{NC}}
\,
\Phi_{\mathrm{B}}
\;.
\label{E503}
\end{eqnarray}
Here
$ \sigma_{{\nu}d}^{\mathrm{NC}}(E) $
is the cross section of the NC process (\ref{E:NC}),
$ E_{\mathrm{th}} $
is the threshold neutrino energy
and
$$
X_{{\nu}d}^{\mathrm{NC}}
\equiv
\int_{E_{\mathrm{th}}} \hskip-3mm
\sigma_{{\nu}d}^{\mathrm{NC}}(E)
X(E)
{\mathrm{d}} E
\simeq
4.7 \times 10^{-43} \,\mathrm{cm}^2
\;.
$$
If
$ \displaystyle
\left\langle
\begin{array}[t]{l}
\displaystyle \sum
\\
\scriptstyle \ell=e,\mu,\tau
\end{array}
\hskip-4mm
\mathrm{P}_{\nu_{e}\to\nu_{\ell}}
\right\rangle_{\hskip-2pt\raisebox{4pt}{\scriptsize{NC}}}
< 1
$
it would mean that
there are transitions of solar
$\nu_e$'s into sterile states.
However,
from Eq.(\ref{E503})
it is clear that we cannot reach any conclusion
on the value of
$ \displaystyle
\left\langle
\begin{array}[t]{l}
\displaystyle \sum
\\
\scriptstyle \ell=e,\mu,\tau
\end{array}
\hskip-4mm
\mathrm{P}_{\nu_{e}\to\nu_{\ell}}
\right\rangle_{\hskip-2pt\raisebox{4pt}{\scriptsize{NC}}}
$
without assumptions about the value of the
total $^8\mathrm{B}$ neutrino flux
$ \Phi_{\mathrm{B}} $.

Let us now consider the ES process (\ref{E:ES}).
In order to separate the NC contribution
to the ES event rate
we must use the information on the flux of solar
$\nu_e$'s on the earth
$ \phi_{\nu_e}(E) $
that can be obtained from the CC process
in SNO.
We have
\begin{equation}
\left\langle
\begin{array}[t]{l}
\displaystyle \sum
\\
\scriptstyle \ell=e,\mu,\tau
\end{array}
\hskip-4mm
\mathrm{P}_{\nu_{e}\to\nu_{\ell}}
\right\rangle_{\hskip-2pt\raisebox{4pt}{\scriptsize{ES}}}
X_{\nu_{\mu}e}
\Phi_{\mathrm{B}}
=
\Sigma^{\mathrm{ES}}
\label{E509}
\end{equation}
with
$$
\begin{array}{l} \displaystyle
\left\langle
\begin{array}[t]{l}
\displaystyle \sum
\\
\scriptstyle \ell=e,\mu,\tau
\end{array}
\hskip-4mm
\mathrm{P}_{\nu_{e}\to\nu_{\ell}}
\right\rangle_{\hskip-2pt\raisebox{4pt}{\scriptsize{ES}}}
\\ \displaystyle
\null \hskip1cm
=
{\displaystyle
\int_{E_{\mathrm{th}}} \hskip-3mm
\sigma_{\nu_{\mu}e}(E)
X(E)
\begin{array}[t]{l}
\displaystyle \sum
\\
\scriptstyle \ell=e,\mu,\tau
\end{array}
\hskip-4mm
\mathrm{P}_{\nu_{e}\to\nu_{\ell}}(E)
\mathrm{d} E
\over\displaystyle
X_{\nu_{\mu}e}
}
\end{array}
$$
and
$$
\Sigma^{\mathrm{ES}}
\equiv
N^{\mathrm{ES}}
-
\int_{{E_{\mathrm{th}}}} \hskip-3mm
\left(
\sigma_{\nu_{e}e}(E)
-
\sigma_{\nu_{\mu}e}(E)
\right)
\phi_{\nu_{e}}(E)
{\mathrm{d}} E
$$
Here $ N^{\mathrm{ES}} $
is the total ES event rate,
$ \sigma_{\nu_\ell e}(E) $
is the cross section of the process
$ \nu_\ell \, e \to \nu_\ell \, e $
(with $\ell=e,\mu$)
and
$$
X_{\nu_{\mu}e}
\equiv
\int_{E_{\mathrm{th}}}
\sigma_{\nu_{\mu}e}(E)
\,
X(E)
\,
{\mathrm{d}} E
\simeq
2 \times 10^{-45} \, \mathrm{cm}^2
$$
for
$ E_{\mathrm{th}} \simeq 6 \, \mathrm{MeV} $
(which corresponds to a
kinetic energy threshold
$ T_{\mathrm{th}} = 4.5 \, \mathrm{MeV} $
for the electrons in the CC process).

Combining the relations (\ref{E503}) and (\ref{E509}),
we obtain
\begin{equation}
{\displaystyle
1
-
\left\langle
\mathrm{P}_{\nu_{e}\to\nu_{s}}
\right\rangle_{\mathrm{ES}}
\over\displaystyle
1
-
\left\langle
\mathrm{P}_{\nu_{e}\to\nu_{s}}
\right\rangle_{\mathrm{NC}}
}
=
{\displaystyle
\Sigma^{\mathrm{ES}}
\,
X_{{\nu}d}^{\mathrm{NC}}
\over\displaystyle
N^{\mathrm{NC}}
\,
X_{\nu_{\mu}e}
}
\equiv
R^{\mathrm{ES}}_{\mathrm{NC}}
\label{E511}
\end{equation}

The ratio
$ R^{\mathrm{ES}}_{\mathrm{NC}} $
is a measurable quantity.
It does not depend on the flux
$ \Phi_{\mathrm{B}} $.
If it will occur that
$ R^{\mathrm{ES}}_{\mathrm{NC}} \not= 1 $
it would mean that there are
transitions of solar $\nu_e$'s
into sterile states.
Of course,
if
$ R^{\mathrm{ES}}_{\mathrm{NC}} = 1 $
it is not possible to reach any
conclusion about the presence of
$ \nu_e \to \nu_s $
transitions
($ R^{\mathrm{ES}}_{\mathrm{NC}} = 1 $
if
$ \mathrm{P}_{\nu_{e}\to\nu_{s}}(E) = 0 $,
but also if
$ \mathrm{P}_{\nu_{e}\to\nu_{s}}(E) = \mbox{constant} $).

In the S-K experiment
the energy spectrum of the recoil electron
will be investigated in detail.
This measurement will allow
to perform other test of the existence
of sterile neutrinos.
We have
\begin{equation}
\left\langle
\begin{array}[t]{l}
\displaystyle \sum
\\
\scriptstyle \ell=e,\mu,\tau
\end{array}
\hskip-4mm
\mathrm{P}_{\nu_{e}\to\nu_{\ell}}
\right\rangle_{\hskip-2pt\raisebox{4pt}{\scriptsize{ES};\scriptsize{T}}}
=
{\displaystyle
\Sigma^{\mathrm{ES}}(T)
\over\displaystyle
X_{\nu_{\mu}e}(T)
\,
\Phi_{\mathrm{B}}
}
\;.
\label{E532}
\end{equation}
Here
\arraycolsep=0cm
\begin{eqnarray}
&&
\Sigma^{\mathrm{ES}}(T)
\equiv
n^{\mathrm{ES}}(T)
-
\label{E534}
\\
&&
\int_{E_{\mathrm{m}}(T)}
\left[
{\displaystyle
\mathrm{d} \sigma_{\nu_{e}e}
\over\displaystyle
\mathrm{d} T
}
(E,T)
-
{\displaystyle
\mathrm{d} \sigma_{\nu_{\mu}e}
\over\displaystyle
\mathrm{d} T
}
(E,T)
\right]
\phi_{\nu_{e}}(E)
\mathrm{d} E
\;,
\nonumber
\end{eqnarray}
is a measurable quantity
($ n^{\mathrm{ES}}(T) $
is the spectrum of the recoil electrons,
$T$ is the electron kinetic energy,
$ \displaystyle
{\displaystyle
\mathrm{d} \sigma_{\nu_{\ell}e}
\over\displaystyle
\mathrm{d} T
}
(E,T)
$
is the differential cross section
of the process
$ \nu_{\ell} e \to \nu_{\ell} e $,
with $\ell=e,\mu,\tau$,
$
E_{\mathrm{m}}(T)
=
{1\over2}
\,
T
\left(
1
+
\sqrt{ 1 + 2 \, m_{e} / T }
\right)
$,
and
$$
X_{\nu_{\mu}e}(T)
\equiv
\int_{E_{\mathrm{m}}(T)}
{\displaystyle
\mathrm{d} \sigma_{\nu_{\mu}e}
\over\displaystyle
\mathrm{d} T
}
(E,T)
X(E)
\mathrm{d} E
$$
is a known function.

{}From Eq.(\ref{E532})
it follows that
\begin{equation}
\left\langle
\begin{array}[t]{l}
\displaystyle \sum
\\
\scriptstyle \ell=e,\mu,\tau
\end{array}
\hskip-4mm
\mathrm{P}_{\nu_{e}\to\nu_{\ell}}
\right\rangle_{\hskip-2pt\raisebox{4pt}{\scriptsize{ES};\scriptsize{T}}}
\le
R^{\mathrm{ES}}(T)
\label{E581}
\end{equation}
with
$$
R^{\mathrm{ES}}(T)
\equiv
{\displaystyle
\Sigma^{\mathrm{ES}}(T)
\over\displaystyle
X_{\nu_{\mu}e}(T)
}
\bigg/
\left(
{\displaystyle
\Sigma^{\mathrm{ES}}(T)
\over\displaystyle
X_{\nu_{\mu}e}(T)
}
\right)_{\mathrm{max}}
$$
Thus,
if the quantity
$ \displaystyle
\Sigma^{\mathrm{ES}}(T)
\big/
X_{\nu_{\mu}e}(T)
$
depends on the energy $T$
it would mean that
solar $\nu_e$'s transform into sterile states.

We have calculated the ratio
$ R^{\mathrm{ES}}(T) $
in a model with
$ \nu_e $--$ \nu_{s} $
mixing and
the values of the mixing parameters
that were obtained by the fit
of the solar neutrino data
\cite{SOLSTE}.
The results of this calculation,
presented in Fig.1,
illustrate the rather strong
$T$-dependence of the ratio
$ R^{\mathrm{ES}}(T) $
in the model.

\begin{figure}[p]
\mbox{\epsfig{file=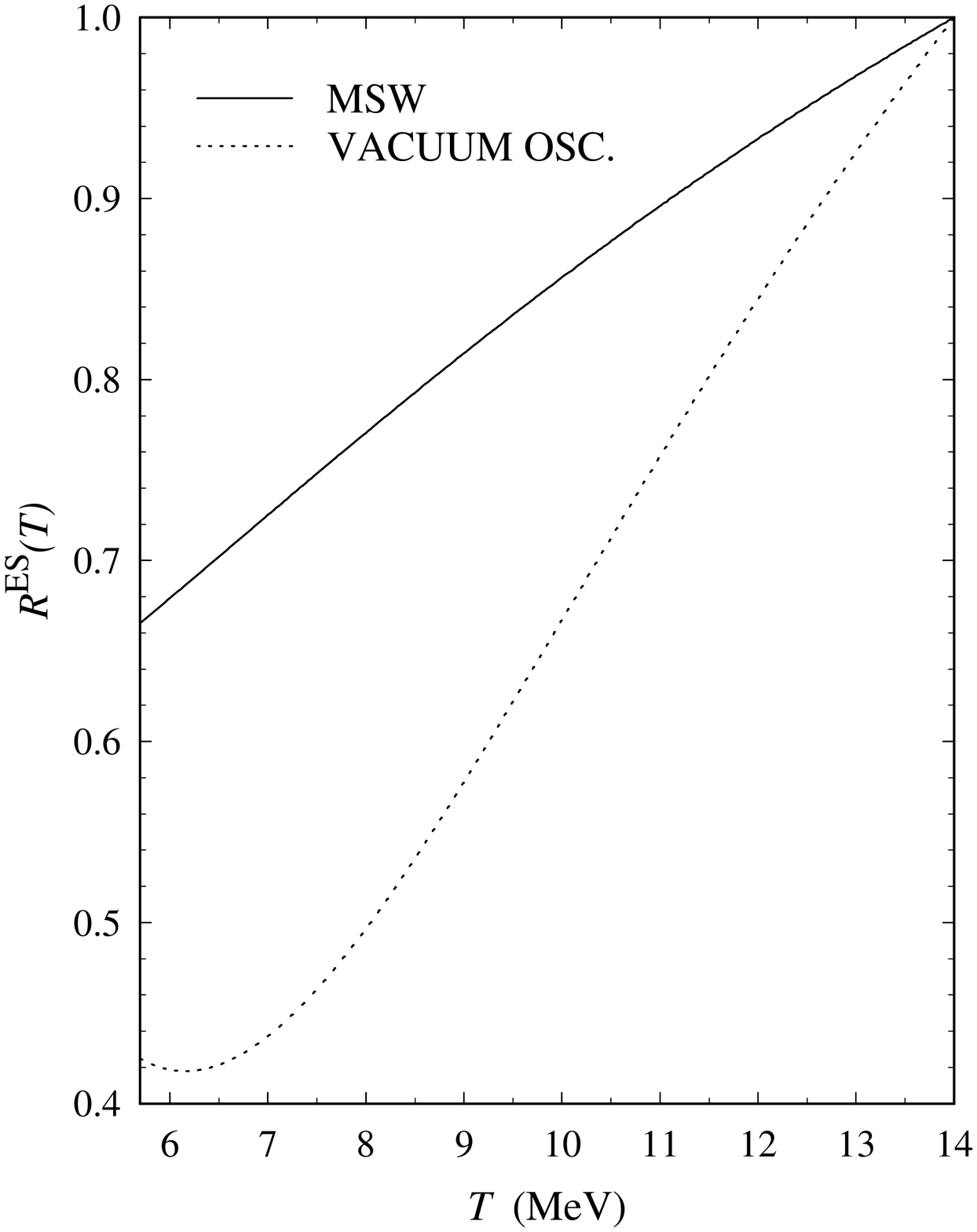,height=0.9\textheight}}
\centerline{Figure 1}
\end{figure}

\end{document}